\documentclass[aps,english,showpacs,twocolumn,pra]{revtex4-1}
\usepackage{amsfonts}
\usepackage{amssymb}
\usepackage{amsmath}
\usepackage{graphicx}
\usepackage{epsfig}
\usepackage{color,xcolor}

\begin{document}

\title{Topology of anti-parity-time-symmetric non-Hermitian
Su-Schrieffer-Heeger model}
\author{H. C. Wu}
\author{L. Jin}
\email{jinliang@nankai.edu.cn}
\author{Z. Song}
\affiliation{School of Physics, Nankai University, Tianjin 300071, China}

\begin{abstract}
We propose an anti-parity-time (anti-$\mathcal{PT}$) symmetric non-Hermitian
Su-Schrieffer-Heeger (SSH) model, where the large non-Hermiticity
constructively creates nontrivial topology and greatly expands the
topological phase. In the anti-$\mathcal{PT}$-symmetric SSH model, the gain
and loss are alternatively arranged in pairs under the inversion symmetry.
The appearance of degenerate point at the center of the Brillouin zone
determines the topological phase transition, while the exceptional points
unaffect the band topology. The large non-Hermiticity leads to unbalanced
wavefunction distribution in the broken anti-$\mathcal{PT}$-symmetric phase
and induces the nontrivial topology. Our findings can be verified through
introducing dissipations in every another two sites of the standard SSH
model even in its trivial phase, where the nontrivial topology is solely
induced by the dissipations.
\end{abstract}

\maketitle

\textit{Introduction.}---The discrete symmetries classify the Hermitian
topological phases into 10 folds \cite{Chiu} and classify the non-Hermitian
topological phases into 38 folds \cite{KawabataPRX}. The many interesting
topological properties of non-Hermitian phases have been reported \cite%
{Ashida}, including the non-Hermitian band theory \cite%
{FuL1,Kawakami,NoriPRB,Wojcik20,KouSP2,KouSP3}, topological insulators \cite%
{ZhuSL,Gilbert1,ZhaiH1,Gilbert2,Budich,Kawabata2,XuZ2,ZhangCW2}, topological
metals \cite{LeeCH3,Bergholtz4,XueP3}, topological semimetals \cite%
{Yoshida1,LMDuan,FanSH2,HZhang2,JHu,Murakami2}, topological invariants \cite%
{DWZhang,WangZ3,Nori1,YiW1,XZZhang1,Hughes2,LeeCH,Hyart,YZhang2}, and
topological edge modes \cite%
{Bardyn,KouSP1,Kohmoto,Kartashov,Malzard2,XueP2,Sone,Lieu2,ZhangXD1}. The
bulk boundary correspondence (BBC) and bulk topological invariant play
important roles in the topological characterization. However, in
non-Hermitian systems with skin effect \cite%
{ZWang1,Jin1,Sato1,Kunst,Ueda1,Szameit3,XueP1,ChenS5,ChenS6,Kawabata3,GongJB2,Longhi4,Budich4}%
, the spectra between open-boundary condition (OBC) and periodic-boundary
condition (PBC) can be dramatically distinct and the conventional BBC is
invalid because of the Aharonov-Bohm effect with imaginary magnetic flux
\cite{Jin1}. To correctly describe spectrum under OBC, the non-Bloch band
theory\ is developed \cite{ZWang1,Murakami,CFang1,Sato5}, the quasimomentum
becomes complex and varies on a generalized Brillouin zone (GBZ). A
universal analytical method to obtain the GBZ is given for one-dimensional
non-Hermitian systems \cite{YangZS1}. In the presence of non-Hermitian skin
effect, the damping becomes unidirectional \cite{WangZ2}. Zener tunneling
becomes chiral at the non-Bloch collapse point \cite{Longhi1}. The relation
between edge modes and bulk topology is formulated using the Green's
function method \cite{Slager}. The non-Hermitian topological systems can be
implemented in many experimental platforms including the passive (active)
photonic crystals of coupled waveguides \cite%
{ChenYF2,ChenYF4,JHou1,Segev,Cerjan2,FuLB,LuPX}, coupled resonators \cite%
{LFeng2,Szameit2,AAlu,Malzard,PengC1,ZhangCW1}, optical lattice \cite%
{LiLH1,Kunst2,ZWZhou}, electronic circuits \cite{Ezawa,Thomale,XXZhang}, and
acoustic lattices \cite{WZhu,BZhang}.

The progresses on the non-Hermitian Su-Schrieffer-Heeger (SSH) models \cite%
{ChenS2,ChenS7,Hughes,Schomerus,Menke,YangWu,XuZ1,Lieu1,LangLJ1,HZhang1,BLiu1,AXChen1,HFWang1,Rocca1,Yuce1,Yuce2,AnJH1,Takane1,Giannini1}%
, Aubry-Andr\'{e}-Harper models \cite%
{Longhi2,JWang1,XuY2,XuY3,ZhangDW3,Longhi3}, and Rice-Mele models \cite%
{YangZS2,WangR,Yuce3} provide fundamental understanding of the non-Hermitian
topological phase of matter. In the non-Hermitian SSH model with asymmetric
couplings, nonzero imaginary magnetic flux \cite{Jin1}, persistent current
\cite{CFang1}, and non-Hermitian skin effect exist. In the parity-time ($%
\mathcal{PT}$) symmetric non-Hermitian SSH model with gain and loss \cite%
{Hughes,Schomerus,ChenS2,Yuce1,Menke,XuZ1,YangWu}, the $\mathcal{PT}$
symmetry prevents nonzero imaginary magnetic flux and ensures the BBC. In
the exact $\mathcal{PT}$ symmetric region with real spectrum, the Berry
phase for each separable band is quantized; in the broken $\mathcal{PT}$
symmetric region with complex spectrum \cite{Christodoulides}, the Berry
phase for each separable band is not quantized \cite{YangWu}. Topological
interface states are experimentally observed in $\mathcal{PT}$ symmetric
non-Hermitian SSH lattices \cite{Poli,Weimann,St-Jean,Parto,LFeng1}.

The anti-$\mathcal{PT}$ symmetry can also protect the validity of BBC. In
this work, we propose an anti-$\mathcal{PT}$-symmetric non-Hermitian SSH
model through alternatively incorporating the balanced gain and loss under
the inversion symmetry in the standard SSH model. The band spectrum becomes
partially complex in the presence of non-Hermiticity, indicating the
thresholdless anti-$\mathcal{PT}$ symmetry breaking. The gain and loss help
creating the nontrivial topology. The topological characterization and the
geometric picture of the topological phases are elaborated. The topological
phase transition occurs when the band gap closes and reopens. The
degenerated topological edge states have zero-energy with net gain and
localized at two lattice boundaries, respectively. Exciting the edge states
enable topological lasing \cite%
{St-Jean,Parto,LFeng1,Kante,Skryabin,Khajavikhan,Iwamoto1}.

\begin{figure}[tb]
\includegraphics[bb=0 0 495 380, width=8.8 cm, clip]{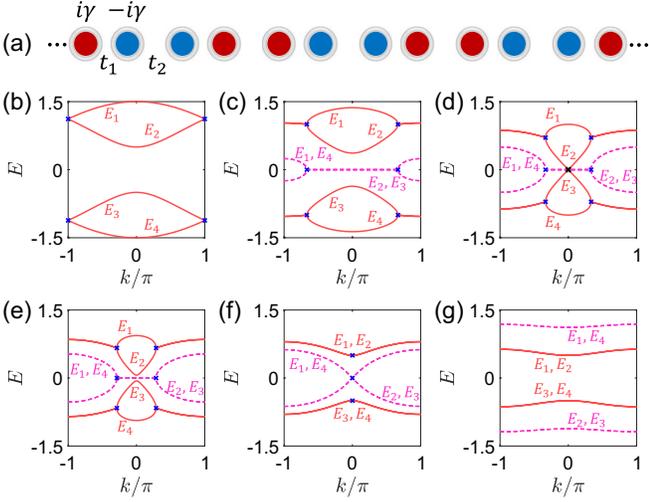}
\caption{(a) Schematic of the anti-$\mathcal{PT}$-symmetric non-Hermitian
lattice of coupled resonator array. The gain (loss) is indicated in red
(blue). The spectra are shown in (b)-(g) with system parameters \ $t_{1}=1$,
$t_{2}=1/2$, and $\protect\gamma =0$ for (b), $\protect\gamma =1/2$ for (c),
$\protect\gamma =\protect\sqrt{3}/2$ for (d), $\protect\gamma =9/10$ for
(e), $\protect\gamma =1$ for (f), $\protect\gamma =3/2$ for (g). The red
lines indicate the real part of energy, and the magenta dashed lines
indicate the imaginary part of energy; where the blue crosses indicate edge
DP for $\protect\gamma =0$ or EPs for $\protect\gamma \neq 0$, the black
cross indicates a central DP in (d).} \label{fig1}
\end{figure}

\textit{Model}.---\label{II}The schematic of the non-Hermitian SSH model is
shown in Fig.~\ref{fig1}(a), which describes a one-dimensional coupled
resonator array. All the resonators have identical resonant frequency. The
staggered distance between the nearest neighbor resonators determines the
lattice couplings $t_{1}$ and $t_{2}$ \cite{Weimann,Poli,
St-Jean,Parto,LFeng1}, which classify two sublattices in the SSH model
\begin{equation*}
H_{0}=\sum_{j}(t_{1}a_{j}^{\dagger }b_{j}+t_{2}b_{j}^{\dagger }a_{j+1}+%
\mathrm{H.c.}),
\end{equation*}%
where $a_{j}^{\dagger }$ ($b_{j}^{\dagger }$) and $a_{j}$ ($b_{j}$) are the
creation and annihilation operators for the sublattice site indexed $j$. To
create the anti-$\mathcal{PT}$ symmetry \cite%
{LYou,Konotop,XFZhu,HJing,Franco,PLu}, the gain and loss are introduced in
the resonators under the inversion symmetry in the form of $\left\{ i\gamma
,-i\gamma ,-i\gamma ,i\gamma \right\} $ in the four-site unit cell
\begin{equation}
H_{1}=i\gamma \sum_{j}(a_{2j-1}^{\dagger }a_{2j-1}-b_{2j-1}^{\dagger
}b_{2j-1}-a_{2j}^{\dagger }a_{2j}+b_{2j}^{\dagger }b_{2j}).
\end{equation}%
The Hamiltonian of the anti-$\mathcal{PT}$-symmetric non-Hermitian SSH model
reads%
\begin{equation}
H=H_{0}+H_{1}.
\end{equation}%
As shown in Fig.~\ref{fig1}(a), the anti-$\mathcal{PT}$-symmetric
non-Hermitian SSH model is invariant under a $\pi $ rotation of the left
(right) non-Hermitian dimer and glide half of the unit cell in the
translational direction \cite{YXZhao16}.

In comparison to the $\mathcal{PT}$-symmetric non-Hermitian SSH model with
the gain and loss $\{i\gamma ,-i\gamma ,i\gamma ,-i\gamma \}$ in the
four-site unit cell, the only difference between the anti-$\mathcal{PT}$%
-symmetric and $\mathcal{PT}$-symmetric SSH models is that the arrangement
of the even number of gain and loss pairs; the anti-$\mathcal{PT}$-symmetric
SSH model has the inversion symmetry, while the $\mathcal{PT}$-symmetric SSH
model does not. Interestingly, the role played by the non-Hermiticity $%
\gamma $ is completely different in these two models. The non-Hermiticity $%
\gamma $ in the anti-$\mathcal{PT}$-symmetric SSH model constructively
create nontrivial topology. The topologically trivial phase changes into the
topologically nontrivial phase as the increasing of non-Hermiticity $\gamma $%
. The nontrivial topology of anti-$\mathcal{PT}$-symmetric SSH model can be
directly verified in many experimental platforms that used to demonstrate
the $\mathcal{PT}$-symmetric SSH model. The topological aspect of the anti-$%
\mathcal{PT}$-symmetric SSH model completely differs from that of the
nonsymmorphic RM model \cite{Fung}.

Applying the Fourier transformation, the Bloch Hamiltonian of the lattice is
obtained as%
\begin{equation}
H_{k}=\left(
\begin{array}{cccc}
i\gamma & t_{1} & 0 & t_{2}e^{-ik} \\
t_{1} & -i\gamma & t_{2} & 0 \\
0 & t_{2} & -i\gamma & t_{1} \\
t_{2}e^{ik} & 0 & t_{1} & i\gamma%
\end{array}%
\right) .  \label{Hk}
\end{equation}%
$H_{k}$ has the anti-$\mathcal{PT}$-symmetry $(\mathcal{PT})H_{k}(\mathcal{PT%
})^{-1}=-H_{k}$\ with $\mathcal{P}=i\sigma _{x}\otimes \sigma _{y}$ and $%
\mathcal{T=K}$ is the complex conjugation operation. $H_{k}$ also has TRS$%
^{\dagger }$ symmetry $\mathcal{C}_{+}H_{k}^{\text{T}}\mathcal{C}%
_{+}^{-1}=H_{-k}$ with $\mathcal{C}_{+}=\sigma _{0}\otimes \sigma _{0}$, PHS$%
^{\dagger }$ symmetry $\mathcal{T}_{-}H_{k}^{\ast }\mathcal{T}%
_{-}^{-1}=-H_{-k}$\ with $\mathcal{T}_{-}=\sigma _{0}\otimes \sigma _{z}$,
and chiral symmetry (pseudo-anti-non-Hermiticity) $\Gamma H_{k}^{\dagger
}\Gamma ^{-1}=-H_{k}$\ with $\Gamma =\mathcal{C}_{+}\mathcal{T}_{-}$, where $%
\sigma _{0}$ and $\sigma _{x,y,z}$ are the two-by-two identical matrix and
Pauli matrix. The system belongs to the BDI$^{\dagger }$ class in the $38$%
-fold topological classifications of non-Hermitian systems and the
topological phase transition of the BDI$^{\dagger }$ class is determined by
the closure of the band gap of the real part of energy bands \cite%
{KawabataPRX}. The interplay between the couplings and the non-Hermiticity
alters the band topology and generates the nontrivial topology; furthermore,
the loss can solely induce the nontrivial topology if we consider a common
gain term $i\gamma $ is removed from $H_{k}$. This greatly simplifies the
verification of the anti-$\mathcal{PT}$-symmetric SSH model in experiments.

In contrast to the $\mathcal{PT}$ symmetry ensures the energy levels to be
conjugate in pairs, the anti-$\mathcal{PT}$ symmetry ensures the energy
levels in pairs with identical imaginary part and opposite real part. The
four energy bands are%
\begin{equation}
E_{\pm ,\pm }=\pm \sqrt{t_{1}^{2}+t_{2}^{2}-\gamma ^{2}\pm 2t_{2}\sqrt{%
t_{1}^{2}\cos ^{2}\left( k/2\right) -\gamma ^{2}}}.
\end{equation}

In the absence of the gain and loss ($\gamma =0$), the lattice is the
Hermitian SSH model. At the topological phase transition point $t_{1}=t_{2}$%
, two bands $\pm 2t_{1}\cos \left( k/2\right) $ of the SSH model are
connected at the degenerate point (DP) $k=\pm \pi $; the four-band spectrum $%
E_{\pm ,\pm }$ of $H_{k}$ can be regarded as the spectrum of the SSH model
folded at $k=\pm \pi /2$ and stretched to the entire Brillouin zone (BZ).
Thus, the central band gap closes at the DP at the center of the BZ $k=0$
and the band folding generates another DP at the edge of the BZ $k=\pm \pi $%
. At $t_{1}\neq t_{2}$, the SSH model is gapped and the central gap is open
as shown in Fig. \ref{fig1}(b); however, the spectrum of $H_{k}$ still has a
DP at the edge of BZ protected by the nonsymmorphic symmetry in the
four-site unit cell of the SSH model \cite{YXZhao16}.

In the presence of the gain and loss ($\gamma \neq 0$), the non-Hermiticity
splits the edge DP into two exceptional points (EPs) associated with the
anti-$\mathcal{PT}$ symmetry breaking [Fig. \ref{fig1}(c)]. As the increase
of the non-Hermiticity, the two EPs gradually move and the complex energy
region expands from the edge to the center of the BZ as shown in Figs. \ref%
{fig1}(d) and \ref{fig1}(e). When $\gamma ^{2}=t_{1}^{2}$, two EPs merge to
one EP at the center of the BZ [Fig. \ref{fig1}(f)] and disappear for $%
\gamma ^{2}>t_{1}^{2}$ [Fig. \ref{fig1}(g)].

The band gap of the central two bands closes at $E=0$ as presented in Figs. %
\ref{fig1}(d), which requires $\left( \gamma^{2}+t_{2}^{2}-t_{1}^{2}\right)
^{2}+4t_{1}^{2}t_{2}^{2}\sin ^{2}\left(k/2\right) =0$. The central two bands
touch at the DP $k=0$ associated with the topological phase transition at
the critical non-Hermiticity%
\begin{equation}
\gamma ^{2}+t_{2}^{2}=t_{1}^{2}.
\end{equation}

\begin{figure}[tb]
\includegraphics[bb=0 0 450 150, width=8.8 cm, clip]{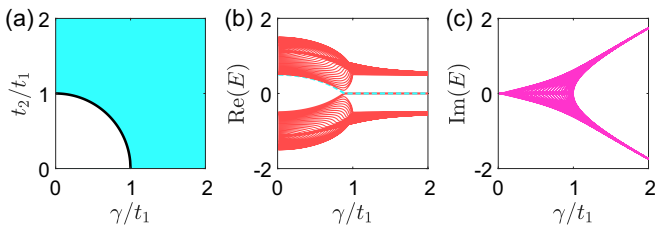}
\caption{(a) Phase diagram for the anti-$\mathcal{PT}$-symmetric
non-Hermitian SSH model. The real part (b) and the imaginary part (c) of the
energy spectrum as a function of the non-Hermiticity; the parameters are $t_{1}=1$ and $t_{2}=1/2$.}
\label{fig2}
\end{figure}

\textit{Phase diagram.---}The anti-$\mathcal{PT}$-symmetric non-Hermitian
SSH model reduces to the SSH model for $\gamma =0$, which has the
topologically nontrivial phase for $t_{2}^{2}>t_{1}^{2}$ and the
topologically trivial phase for $t_{2}^{2}<t_{1}^{2}$. However, the
situation changes in the presence of non-Hermiticity as shown in the phase
diagram Fig. \ref{fig2}(a). The non-Hermiticity creates the nontrivial
topology and the topological region expands in the anti-$\mathcal{PT}$%
-symmetric non-Hermitian SSH model, where $\gamma ^{2}+t_{2}^{2}<t_{1}^{2}$
is the topologically trivial phase and $\gamma ^{2}+t_{2}^{2}>t_{1}^{2}$ is
the topologically nontrivial phase. The nontrivial topology of the anti-$%
\mathcal{PT}$-symmetric SSH model can be solely created by the
non-Hermiticity because large non-Hermiticity induces unbalanced
distributions of the wavefunction probability. The non-Hermiticity generates
nontrivial topology in the uniform chain at $t_{1}^{2}=t_{2}^{2}$ \cite%
{KTakata18} and even in the trivial phase of the Hermitian SSH model at $%
t_{2}^{2}<t_{1}^{2}$.

\begin{figure}[tb]
\includegraphics[bb=0 0 505 315, width=8.8 cm, clip]{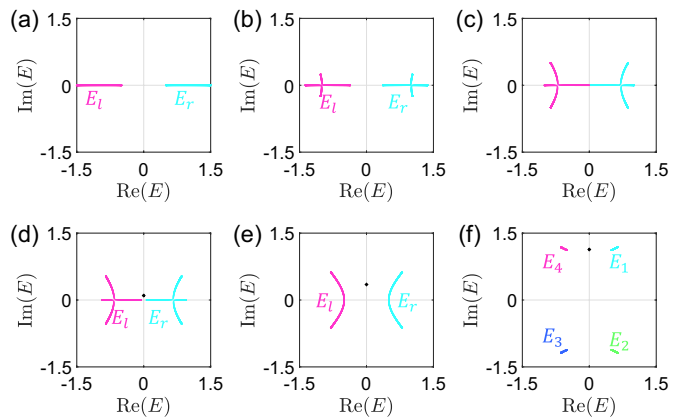}
\caption{Energy bands in the complex energy plane under OBC. The parameters
in (a)-(f) are similar as those in Figs.~\protect\ref{fig1}(b)-\protect\ref{fig1}(g). The black stars indicate a pair of edge states for topological
nontrivial phases as shown in (d)-(f).} \label{fig3}
\end{figure}

The real part and imaginary part of the energy bands under OBC as a function
of the non-Hermiticity are shown in Figs. \ref{fig2}(b) and \ref{fig2}(c),
respectively. The anti-$\mathcal{PT}$-symmetric non-Hermitian SSH lattice
under OBC has one pair of edge states in the topologically nontrivial phase.
To further elucidate the band structure, the energy bands in the complex
energy plane for the non-Hermitian SSH lattice under OBC are plotted as
shown in Figs. \ref{fig3}(a)-\ref{fig3}(f); the corresponding PBC spectra
are shown in Figs. \ref{fig1}(b)-\ref{fig1}(g). For $\gamma ^{2}\leqslant
t_{1}^{2}$ and $\gamma ^{2}+t_{2}^{2}\neq t_{1}^{2}$, the real part of
energy bands is gapped (the central gap is open). However, the energy bands $%
E_{1}$ and $E_{2}$ ($E_{3}$ and $E_{4}$) are inseparable \cite{FuL1} because
of the existence of edge DP or EPs (blue crosses) as shown in Figs. \ref%
{fig1}(b)-\ref{fig1}(c) and \ref{fig1}(e)-\ref{fig1}(f); in this sense, the
four energy bands can be regarded as two energy bands $E_{r}$ (cyan) and $%
E_{l}$ (magenta) according to the real part of energy bands as shown in
Figs. \ref{fig3}(a)-\ref{fig3}(b) and \ref{fig3}(d)-\ref{fig3}(e). For $%
\gamma ^{2}+t_{2}^{2}=t_{1}^{2}$ (topological phase transition), the real
gap is closed (the central gap is closed) and the two energy bands $E_{r}$
and $E_{l}$ become single band [Fig. \ref{fig3}(c)]. For $\gamma
^{2}>t_{1}^{2}$, the EP disappears between $E_{1}$ and $E_{2}$ ($E_{3}$ and $%
E_{4}$) and the four energy bands are separated [Fig. \ref{fig3}(f)]. In the
topologically nontrivial region $\gamma ^{2}+t_{2}^{2}>t_{1}^{2}$, there
exist one pair of degenerated zero modes (black stars) with gain for the
lattice under OBC as shown in Figs. \ref{fig3}(d)-\ref{fig3}(f).

\textit{Edge state.---}The inversion symmetry ensures the validity of
conventional BBC in the anti-$\mathcal{PT}$-symmetric non-Hermitian lattice
\cite{Jin1}. The lattice under OBC supports two degenerate edge states
localized at the left and right boundaries of the lattice, respectively. The
eigenvalues of edge states are imaginary%
\begin{equation}
\varepsilon =i\left\{ \lambda +\left[ 4\gamma ^{2}-3(t_{2}^{2}+t_{1}^{2})%
\right] /\lambda -\gamma \right\} /3,
\end{equation}%
where $\lambda =\sqrt[3]{\alpha +\sqrt{\beta }}$ with $\alpha
=(18t_{2}^{2}+8\gamma ^{2}-9t_{1}^{2})\gamma $ and $\beta =27\{\left[
4(2t_{2}^{2}+4\gamma ^{2}-5t_{1}^{2})t_{2}^{2}-t_{1}^{4}\right] \gamma
^{2}+\left( t_{1}^{2}+t_{2}^{2}\right) ^{3}\}$. The left edge state
localizes at the left boundary of the lattice. Without loss of generality,
the wave functions in the $j$-th unit cell of the left edge state $%
\left\vert \psi _{L}\right\rangle $ can be expressed as $\chi ^{j-1}\left\{
1,\left( \varepsilon -i\gamma \right) /t_{1},-\chi t_{2}/t_{1},0\right\} $
with $\chi =(i\gamma -\varepsilon )/\left( i\gamma +\varepsilon \right) $.
For $\gamma =0$, the edge states reduce to zero modes with $\varepsilon =0$
and the wave functions in the $j$-th unit cell as $\left( t_{1}/t_{2}\right)
^{j-1}\left\{ 1,0,-t_{1}/t_{2},0\right\} $. The right edge state $\left\vert
\psi _{R}\right\rangle $ is the mirror reflection of the left edge state $%
\left\vert \psi _{L}\right\rangle $. The edge states have net gain rate and
are useful for topological lasing.

\textit{Geometric picture of band topology.---}The topology of the anti-$%
\mathcal{PT}$-symmetric non-Hermitian SSH model relates to the geometry of
the Bloch Hamiltonian winding around the degeneracy points in a two
dimensional parameter space, we show how the non-Hermiticity creates the
nontrivial topology. For the Bloch Hamiltonian Eq. (\ref{Hk}), we replace $%
e^{ik}$ by $h_{x}+ih_{y}$ to create a two dimensional parameter space $%
(h_{x},h_{y})$
\begin{equation}
H(h_{x},h_{y})=\left(
\begin{array}{cccc}
i\gamma & t_{1} & 0 & t_{2}(h_{x}-ih_{y}) \\
t_{1} & -i\gamma & t_{2} & 0 \\
0 & t_{2} & -i\gamma & t_{1} \\
t_{2}(h_{x}+ih_{y}) & 0 & t_{1} & i\gamma%
\end{array}%
\right) ,
\end{equation}%
where the $k$-dependent Bloch Hamiltonian Eq. (\ref{Hk}) corresponds to a
unit circle $h_{x}^{2}+h_{y}^{2}=1$ in the two dimensional parameter space $%
(h_{x},h_{y})$.

\begin{figure}[t]
\includegraphics[bb=0 0 495 630, width=8.8 cm, clip]{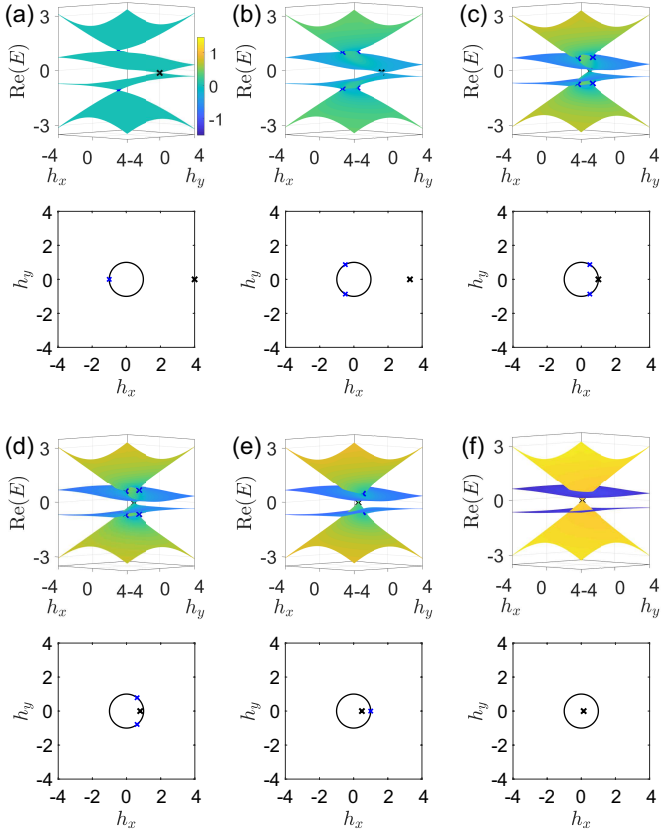}
\caption{Complex energy bands and geometric picture of the topology in the
two dimensional parameter space $(h_{x},h_{y})$ for the anti-$\mathcal{PT}$-symmetric non-Hermitian SSH model. The upper panels depict the complex
energy bands, where the value indicates the real part and the color
indicates the imaginary part of the energy; the lower panels depict the
geometric picture of the topology. The unit circle indicates the anti-$\mathcal{PT}$-symmetric non-Hermitian SSH model, the blue crosses indicate
the edge DP or EPs, and the black crosses indicate the central DP. The
parameters are (a) $\protect\gamma =0$, (b) $\protect\gamma =1/2$, (c) $\protect\gamma =\protect\sqrt{3}/2$, (d) $\protect\gamma =9/10$, (e) $\protect\gamma =1$, (f) $\protect\gamma =3/2$; other parameters are $t_{1}=1$
and $t_{2}=1/2$.} \label{fig4}
\end{figure}

Figure \ref{fig4} depicts the complex energy bands\ extended to the two
dimensional parameter space $(h_{x},h_{y})$ for the anti-$\mathcal{PT}$%
-symmetric non-Hermitian SSH model at fixed parameters $t_{1}=1$ and $%
t_{2}=1/2$ for different non-Hermiticity $\gamma $. The edge DP (blue cross)
on the unit circle at $\gamma =0$ splits into two EPs on the unit circle at
nonzero non-Hermiticity $\gamma <1$; and the two EPs are symmetrically
distributed about $h_{y}=0$ because the EPs are symmetrically distributed
about $k=0$ in the BZ [{see Figs. \ref{fig1}(c)-\ref{fig1}(e)]. As the
non-Hermiticity increases, the two EPs move on the unit circle from $\left(
h_{x},h_{y}\right) =\left( -1,0\right) $ to $\left( h_{x},h_{y}\right)
=\left( 1,0\right) $; at $\gamma =1$, the two EPs merge into single EP at $%
\left( h_{x},h_{y}\right) =\left( 1,0\right) $; and the EP vanishes for $%
\gamma >1$. The edge DP or the EPs remain on the unit circle, the topology
is fully determined by the central DP (black cross). From the unit circle
winding around the central DP, we can observe how the nontrivial topology is
created by the non-Hermiticity $\gamma $. For $\gamma =0$ in Fig. \ref{fig4}%
(a), the central DP is outside the unit circle; thus, the system is in the
topologically trivial phase. The nonzero non-Hermiticity $\gamma $ moves the
central DP along $h_{y}=0$ towards the negative $h_{x}$ direction in the
parameter space. For $\gamma =1/2$ in Fig. \ref{fig4}(b), the central DP
moves to $(h_{x},h_{y})=(3.284,0)$, and the system enters the white region
of the phase diagram as shown in Fig. \ref{fig2}(a). For $\gamma =\sqrt{3}/2$
in Fig. \ref{fig4}(c), the central DP moves to $(h_{x},h_{y})=(1,0)$ and
locates on the unit circle; the system is at the boundary of the white and
cyan regions. For $\gamma =9/10$ in Fig. \ref{fig4}(d), the central DP is
enclosed in the unit circle, the topology of the system changes and the
system enters the cyan region. For $\gamma =1$ in Fig. \ref{fig4}(e), the
central DP keeps inside the unit circle; the system is in the nontrivial
phase for the nonzero }$t_{2}$, but the energy bands are not completely
separated{. For $\gamma =3/2$ in Fig. \ref{fig4}(f), the central DP is still
inside the unit circle; in this situation, the four bands are completely
separated and the system is in the nontrivial phase.}

\begin{figure}[tb]
\includegraphics[bb=0 0 470 155, width=8.8 cm, clip]{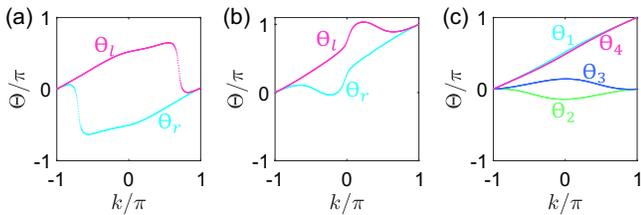}
\caption{Zak
phase and \emph{partial} global Zak phase. The parameters are $t_{1}=1$,
$t_{2}=1/2$, and $\protect\gamma =1/2$ for (a), $\protect\gamma =9/10$ for
(b), $\protect\gamma =3/2$ for (c).} \label{fig5}
\end{figure}

\textit{Zak phase and }\emph{partial} \textit{global Zak phase}.---When the
four complex energy bands are separated at $\gamma ^{2}>t_{1}^{2}$, each
energy band is associated with a Zak phase
\begin{equation}
\Theta _{n}=i\oint \mathrm{d}k\left\langle \varphi _{n}\right\vert \partial
_{k}\left\vert \psi _{n}\right\rangle .
\end{equation}%
In the definition of $\Theta _{n}$, $\left\vert \varphi _{n}\right\rangle $
is the left eigenstate and $\left\vert \psi _{n}\right\rangle $ is the right
eigenstate, $H_{k}\left\vert \psi _{n}\right\rangle =E_{n}\left\vert \psi
_{n}\right\rangle $ and $H_{k}^{\dagger }\left\vert \varphi
_{n}\right\rangle =E_{n}^{\ast }\left\vert \varphi _{n}\right\rangle $,
where the subscript $n$ is the band index. $E_{1}$ denotes the band with
positive real and imaginary energy, $E_{2}$ denotes the band with positive
real and negative imaginary energy, $E_{3}$ denotes the band with negative
real and imaginary energy, and $E_{4}$ denotes the band with negative real
and positive imaginary energy as shown in Figs. \ref{fig3}(f) and Fig. \ref%
{fig1}(g). Their wavefunctions are $\left\vert \psi _{1}\right\rangle $, $%
\left\vert \psi _{2}\right\rangle $, $\left\vert \psi _{3}\right\rangle $,
and $\left\vert \psi _{4}\right\rangle $, respectively. The system has the
inversion symmetry, which ensures that the Zak phase for each separated
energy band is an integer of $\pi $. Thus, the Zak phase is used for
topological characterization. In Fig. \ref{fig5}(c), the Zak phases for the
bands $E_{1}$ and $E_{4}$ are $\pi $ and the Zak phases for the bands $E_{2}$
and $E_{3}$ are $0$;\ which are consistent with the geometric picture in
Fig. \ref{fig4}(f), the central DP belongs to energy bands $E_{1}$\ and $%
E_{4}$, and predicts the existence of one pair of the topological zero modes
with gain for the system under OBC.

For the energy bands embedded with EPs ($\gamma ^{2}\leqslant t_{1}^{2}$ and
$\gamma ^{2}+t_{2}^{2}\neq t_{1}^{2}$), there are only two energy bands $%
E_{r}$ and $E_{l}$. The two-state coalescence EP2$_{12}$ exists only in the
energy bands $E_{1}$ and $E_{2}$, and the two-state coalescence EP2$_{34}$
exists only in the energy bands $E_{3}$ and $E_{4}$. In this sense, we
define two \emph{partial} global Zak phase
\begin{equation}
\Theta _{r}=\Theta _{1}+\Theta _{2};\Theta _{l}=\Theta _{3}+\Theta _{4}.
\end{equation}%
In the calculation of the \emph{partial} global Zak phase, the momentum
ranges $[k_{\text{EP}}-\Delta k,k_{\text{EP}}+\Delta k]$\ are removed
because that the coalesced wavefunctions are self-orthogonal at the EPs \cite%
{KTakata18}, where $\Delta k$ is an infinite small positive real number. The
\emph{partial} global Zak phase is valid for the topological
characterization.

For $\gamma ^{2}\leqslant t_{1}^{2}$ and $\gamma ^{2}+t_{2}^{2}<t_{1}^{2}$,
both the \emph{partial} global Zak phase $\Theta _{r}$ and $\Theta _{l}$ are
$0$ as shown in Fig. \ref{fig5}(a). This indicates the phase is
topologically trivial without any edge state under OBC. However, for $\gamma
^{2}\leqslant t_{1}^{2}$ and $\gamma ^{2}+t_{2}^{2}>t_{1}^{2}$, both the
\emph{partial} global Zak phase $\Theta _{r}$ and $\Theta _{l}$ are $\pi $
as shown in Fig. \ref{fig5}(b). This indicates the phase is topologically
nontrivial and one pair of topological edge states appear under OBC.

\textit{Conclusion.}---We propose the anti-$\mathcal{PT}$-symmetric
non-Hermitian SSH model as a prototypical anti-$\mathcal{PT}$-symmetric
topological lattice. The gain and loss are alternatively introduced in pairs
in the standard SSH model through holding the inversion symmetry. The
inversion symmetric gain and loss result in the thresholdless breaking of
anti-$\mathcal{PT}$ symmetry and the energy spectrum is partially or fully
complex. We provide novel insights on the roles played by the anti-$\mathcal{%
PT}$-symmetry and non-Hermiticity in the topological phases. The large
non-Hermiticity constructively creates the nontrivial topology and greatly
expands the topologically nontrivial region of the SSH model. The
topological edge states localized at two boundaries of the lattice are
degenerate and suitable for topological lasing. Besides, the dissipation can
solely induce the nontrivial topology. In comparison to the $\mathcal{PT}$%
-symmetric non-Hermitian SSH model, only the arrangement of gain and loss in
the anti-$\mathcal{PT}$-symmetric non-Hermitian SSH model is different; the
proposed anti-$\mathcal{PT}$-symmetric non-Hermitian SSH model can be easily
implemented in the microring resonator arrays, coupled optical waveguides,
photonic crystals, electronic circuits, and acoustic lattices~\cite%
{ChenYF2,ChenYF4,JHou1,Segev,Cerjan2,LFeng2,Szameit2,AAlu,Malzard,PengC1,ZhangCW1,LiLH1,Kunst2,Ezawa,Thomale,XXZhang,WZhu,BZhang,FuLB,LuPX,ZWZhou}%
.

This work was supported by National Natural Science Foundation of China
(Grants No.~11975128 and No.~11874225).

\end{document}